\documentclass{rnaastex}
\pdfoutput=1 %for arXiv submission
\usepackage{amsmath,amstext}
\usepackage[T1]{fontenc}
\usepackage{apjfonts} 
\usepackage[figure,figure*]{hypcap}
\usepackage{color}
\usepackage{booktabs}

\newcommand{\unit}[1]{\ensuremath{\, \mathrm{#1}}}

\usepackage{hyperref}
\usepackage{breakurl}
\usepackage{url}
\begin{document}
\title{A Candidate Transit Event around Proxima Centauri}
\email{yxl5559@psu.edu}

%% The \author command can take an optional ORCID.
\author{Yiting Li}
\affiliation{Center for Exoplanets \& Habitable Worlds, Department of Astronomy \& Astrophysics, The Pennsylvania State University}

\author{Gudmundur Stefansson}
\affiliation{Center for Exoplanets \& Habitable Worlds, Department of Astronomy \& Astrophysics, The Pennsylvania State University}
%\affiliation{Center for Exoplanets \& Habitable Worlds, University Park, PA 16802, USA}
\affiliation{NASA Earth and Space Science Fellow}

\author{Paul Robertson}
\affiliation{NASA Sagan Fellow}
\affiliation{Center for Exoplanets \& Habitable Worlds, Department of Astronomy \& Astrophysics, The Pennsylvania State University}
%\affiliation{Center for Exoplanets \& Habitable Worlds, University Park, PA 16802, USA}

\author{Andrew Monson}
\affiliation{Center for Exoplanets \& Habitable Worlds, Department of Astronomy \& Astrophysics, The Pennsylvania State University}
%\affiliation{Center for Exoplanets \& Habitable Worlds, University Park, PA 16802, USA}

\author{Caleb Ca\~{n}as}
\affiliation{Center for Exoplanets \& Habitable Worlds, Department of Astronomy \& Astrophysics, The Pennsylvania State University}
%\affiliation{Center for Exoplanets \& Habitable Worlds, University Park, PA 16802, USA}

\author{Suvrath Mahadevan}
\affiliation{Center for Exoplanets \& Habitable Worlds, Department of Astronomy \& Astrophysics, The Pennsylvania State University}
%\affiliation{Penn State Astrobiology Research Center, University Park, PA 16802, USA}
%\affiliation{Center for Exoplanets \& Habitable Worlds, University Park, PA 16802, USA}

%% Note that RNAAS manuscripts DO NOT have abstracts.
%% Start the main body of the article. If no sections in the 
%% research note leave the \section call blank to make the title.

\vspace*{+5mm}
\section{}
Proxima Centauri is the tertiary companion to the $\alpha$\,Cen binary, and the nearest star to the Sun. Recent results indicate Proxima represents a compact analog to the Solar system.  The Pale Red Dot radial velocity (RV) campaign \citep[PRD;][]{Proxima2016} discovered the star hosts a low-mass exoplanet in a temperate orbit. Radio observations \citep{ALMA2017} also revealed at least two dust belts around Proxima.

Prospects for followup of a planet transiting the nearest star prompted efforts to detect a transit of Proxima b \citep{Kipping2017,liu17}. These attempts were complicated by high activity-induced variability \citep{MOST2016}, so while candidate transits of b have been identified, they are inconclusive.

Towards the end of the PRD campaign, we began a blind search for transits of Proxima in anticipation of the likely announcement of exoplanets in the system. This search yielded one candidate transit inconsistent with the transit window of Proxima b. If this event is due to an exoplanet, its depth and duration are indicative of a second low-mass exoplanet orbiting Proxima.

We observed Proxima over 23 nights from August 17--September 27, 2016 for $\sim$9 hours per night, using a robotic 30cm telescope \citep{monson17} at Las Campanas Observatory. The airmass of Proxima varied between 1 and 3. We defocused the telescope to give a 10$^{\arcsec}$ PSF and set an average exposure time of 15s (readout time 12s). We performed differential aperture photometry using AstroImageJ \citep{collins2017}.
Our unbinned precision was $\sim$1\%, except for three nights where the precision was 0.5\%. A signal was identified on August 25, when Proxima was relatively quiet, and we see no correlation with atmosphere/instrument parameters. Our data are available online\footnote{\url{https://github.com/Yiting206265/Proxima-Photometry}}.

\begin{figure}[h]
	\begin{center}
		\includegraphics[width=\columnwidth]{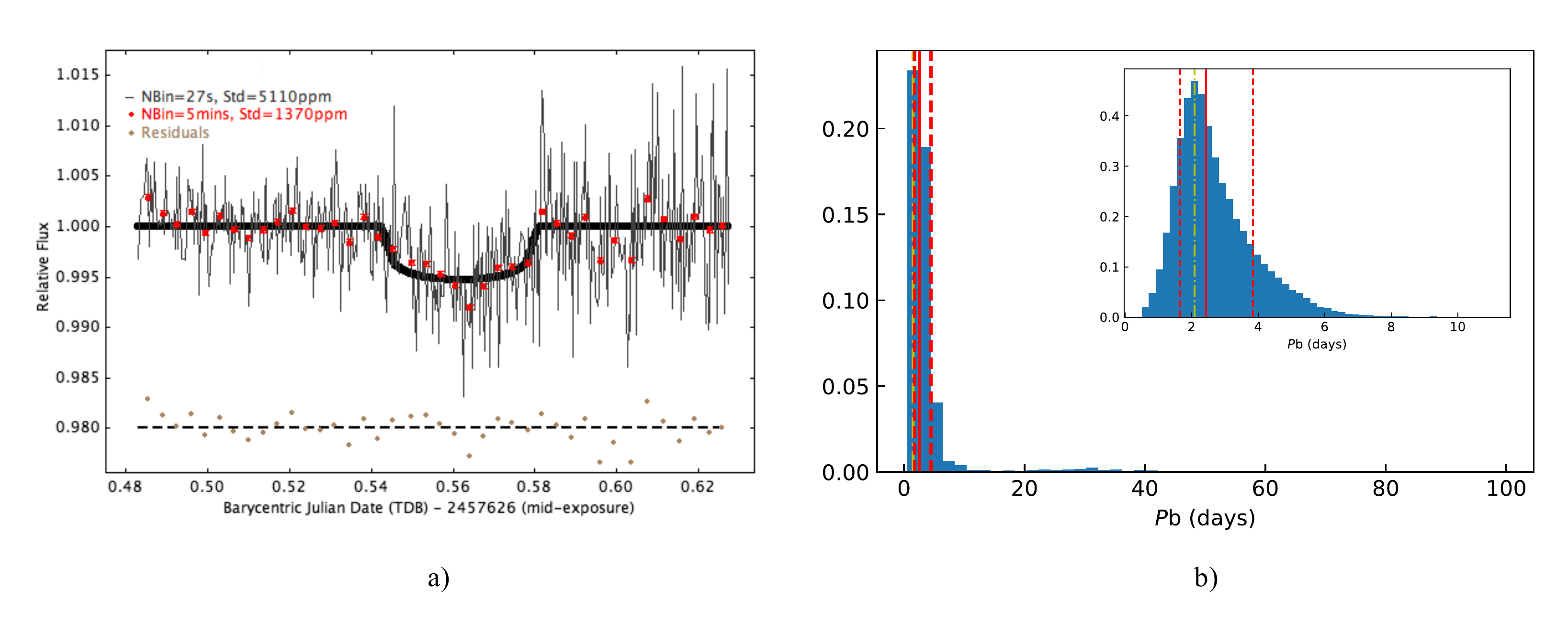}
	\end{center}
    \vspace*{-5mm}
	\caption{\textit{a)} August 25 light curve and the best-fit transit model. Residuals are shown below. \textit{b)} Predicted period distribution.}
	\label{fig:transit}
\end{figure}

We analyzed the candidate transit using \texttt{pyaneti} \citep{pyaneti}, primarily to constrain the period distribution of the transiting body. Assuming zero eccentricity, we put a Gaussian prior $\rho_* = 47.0^{+4.5}_{-4.0} \unit{g/cm^3}$ on the stellar density using literature values from \citep{Proxima2016}, and a uniform prior on the period from $0.5-100 \unit{days}$. \texttt{pyaneti} is capable of fitting RVs and transits, but we only fit the light curve because the event is incompatible with a transit of Proxima b, occurring $>3$ days before the transit time predicted by the orbit from \citet{Proxima2016}. Our model (Figure \ref{fig:transit}) indicates a transit depth of 5 mmag (yielding $R_p \sim 1 R_{\oplus}$) 
, a mid-transit time of $2457626.5635537^{+ 0.0015813}_{- 0.0023548}$ $\rm (BJD_{TDB})$, and a duration of $1 \unit{hour}$. 

%-------------------------------------------------------------
%-------------------------------------------------------------
%-------------------------------------------------------------
%-------------------------------------------------------------

While we are confident this event was not caused by weather or instrumental systematics, its origin is otherwise ambiguous. Stellar activity is a persistent concern, but we expect lower variability in the $I_{C}$ band, and the shape of the event is more consistent with a transit. If it is a transit, the presence of dust belts suggests the transiting body could be a collection of debris instead of an exoplanet. If we attribute this event to an exoplanet, our model prefers a $2-4 \unit{day}$ orbit (Figure \ref{fig:transit}). The available RVs place strong limits on planets in those orbits; any companion more massive than $\sim0.4 \unit{M_{\oplus}}$ should be visible. Thus, if the planet has a radius $\sim 1 \unit{R_{\oplus}}$, it must have a mass consistent with the lowest-density small planets \citep[e.g.][]{lissauer13} to have eluded RV detection. The alternative is that its orbit is external to Proxima b, but that renders the likelihood that we would have observed its transit in our search improbably small. 

Our candidate transit therefore offers tantalizing, inconclusive evidence for a small body transiting Proxima Centauri. We urge renewed efforts to identify additional transits.

% ---------------------------
% Bibliography
% ---------------------------

%\pagebreak
\bibliographystyle{yahapj}
\bibliography{RNAAS_Proxima_Cen}

\acknowledgments

We acknowledge support from the Penn State CEHW, NESSF, the Sagan Fellowship Program, and NSF grants AST-1006676, 1126413, 1310885, 1517592.  We thank Barry Madore for initiating the TMMT Project.

%Software:  AstroImageJ \citep{collins2017}, \texttt{pyaneti}  \citep{pyaneti}.

\end{document}